\documentclass[10pt,sigconf,nonacm,screen]{acmart}

\pdfoutput=1
\usepackage{pdfpages}

\usepackage[all]{nowidow}
\usepackage{xcolor}
\definecolor{lightblue}{rgb}{0.80, 0.8, 1.00}
\usepackage{subcaption}

\usepackage{longtable}
\usepackage{tikz}
\usetikzlibrary{shapes.geometric, arrows, shadows, positioning}
\usepackage{pgfplots}
\pgfplotsset{compat=1.16}

\usepackage[capitalise,noabbrev]{cleveref}
\crefformat{section}{\S#2#1#3} 
\crefformat{subsection}{\S#2#1#3}
\crefformat{subsubsection}{\S#2#1#3}
\crefrangeformat{section}{\S#3#1#4 to~\S#5#2#6}
\crefrangeformat{subsection}{\S#3#1#4 to~\S#5#2#6}
\crefrangeformat{subsubsection}{\S#3#1#4 to~\S#5#2#6}

\begin{document}
\includepdf[pages=-,lastpage=1, scale=1.0]{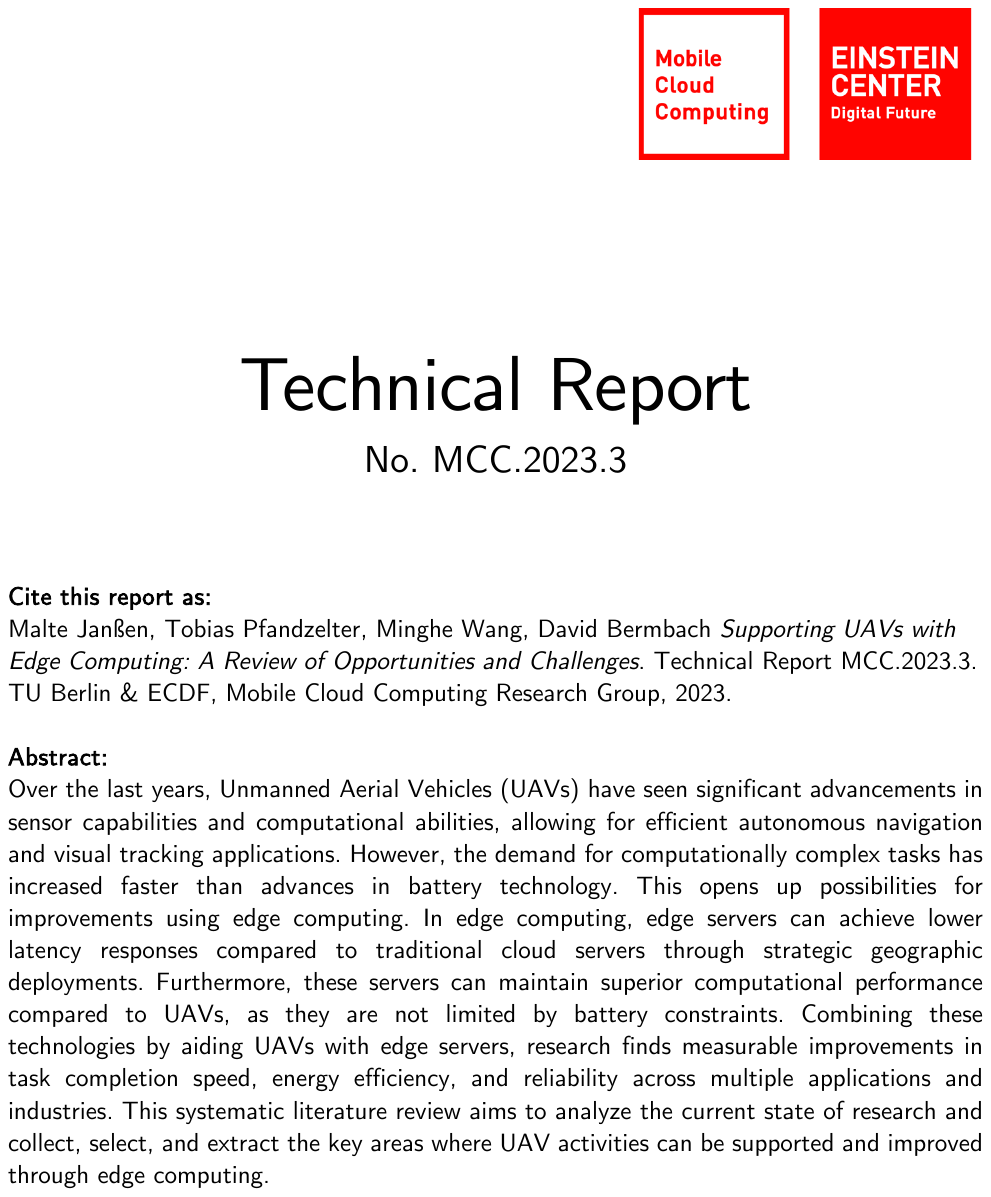}

\author{Malte Jan\ss{}en}
\affiliation{%
    \institution{TU Berlin \& ECDF}
    \department{Mobile Cloud Computing Research Group}
    \city{Berlin}
    \country{Germany}}
\email{mja@mcc.tu-berlin.de}

\author{Tobias Pfandzelter}
\affiliation{%
    \institution{TU Berlin \& ECDF}
    \department{Mobile Cloud Computing Research Group}
    \city{Berlin}
    \country{Germany}}
\email{tp@mcc.tu-berlin.de}

\author{Minghe Wang}
\affiliation{%
    \institution{TU Berlin \& ECDF}
    \department{Mobile Cloud Computing Research Group}
    \city{Berlin}
    \country{Germany}}
\email{mw@mcc.tu-berlin.de}

\author{David Bermbach}
\affiliation{%
    \institution{TU Berlin \& ECDF}
    \department{Mobile Cloud Computing Research Group}
    \city{Berlin}
    \country{Germany}}
\email{db@mcc.tu-berlin.de}

\title{Supporting UAVs with Edge Computing: A Review of Opportunities and Challenges}


\begin{abstract}
    Over the last years, Unmanned Aerial Vehicles (UAVs) have seen significant advancements in sensor capabilities and computational abilities, allowing for efficient autonomous navigation and visual tracking applications.
    However, the demand for computationally complex tasks has increased faster than advances in battery technology.
    This opens up possibilities for improvements using edge computing.
    In edge computing, edge servers can achieve lower latency responses compared to traditional cloud servers through strategic geographic deployments.
    Furthermore, these servers can maintain superior computational performance compared to UAVs, as they are not limited by battery constraints.
    Combining these technologies by aiding UAVs with edge servers, research finds measurable improvements in task completion speed, energy efficiency, and reliability across multiple applications and industries.
    This systematic literature review aims to analyze the current state of research and collect, select, and extract the key areas where UAV activities can be supported and improved through edge computing.
\end{abstract}

\maketitle

\section{Introduction}
\label{sec:introduction}

Edge computing and unmanned aerial vehicles (UAVs) are two areas of technology that have seen notable development in recent years~\cite{Khan2019, Shakhatreh2019}.
This article aims to provide a systematic literature review of the key benefits and limitations of integrating edge computing within UAVs, according to existing literature.

\textbf{Edge Computing} is a distributed computing paradigm that brings computation and data storage closer to the location where it is needed~\cite{Shi2016,paper_bermbach2017_fog_vision}.
Traditionally, cloud servers are located in centralized data centers, often far away from the devices that require their services.
This can lead to increased latency, especially for applications that require real-time data processing.
Edge computing aims to solve these issues by deploying edge servers in close proximity to the devices, splitting up the centralized cloud into a distributed network of edge servers.
This computing approach aims to improve response times and save bandwidth, thereby allowing real-time data processing and enhancing user experience in a wide range of applications.

\textbf{Unmanned Aerial Vehicles (UAVs)}, commonly also referred to as `drones', are aircraft without a human pilot onboard.
These vehicles can be remotely controlled by an operator on the ground or can fly autonomously using pre-programmed flight plans or more complex automations.
These vehicles are battery-powered and can carry a variety of sensors, such as cameras or Lidar sensors.
Because of their small size and maneuverability, UAVs are often used in applications where traditional aircraft are not feasible because of economic or safety concerns.

The intersection of edge computing and UAVs allows for improvements in different directions.
For one, activities carried out by UAVs can be aided by edge servers, e.g.
to improve speed, latency or reliability of these activities.
Another possible direction is to use UAVs as part of an existing edge computing system, e.g.
by covering areas not reachable by traditional, stationary deployments.
This systematic literature review will focus on the former, as there is currently less research on possible improvements.
Utilizing edge computing presents numerous potential benefits, from enhanced real-time data processing capabilities to improved operational efficiency.
However, it also poses several challenges and limitations that need to be addressed.
This systematic literature review seeks to understand and outline these key benefits and limitations as presented in the existing literature.

The research question guiding this review is: ``What are the key benefits and limitations of aiding UAVs using edge computing according to existing literature?''
This literature review is structured as follows: \cref{sec:methodology} outlines the research methodology, including the search strategy, study selection, and data extraction and analysis.
\cref{sec:results} presents the results of the literature review, including the search outcomes and common themes.
\cref{sec:discussion} discusses the findings and highlights research gaps.
\cref{sec:limitations} outlines the limitations of this review.
Finally, \cref{sec:conclusion} concludes the review.

\section{Methodology}
\label{sec:methodology}

This section outlines the research method used to identify, select, and evaluate the relevant literature, as well as the process followed for data extraction and synthesis.

\subsection{Search Strategy}

Relevant literature can be found in various databases, including Google Scholar, SpringerLink, ScienceDirect, IEEE Xplore, and ACM Digital Library.
All of these databases offer extensive collections of computer science and technology literature.

Following the standard guidelines for systematic literature reviews as established by~\citet{kitchenham2007}, the search terms were identified on the basis of the research question, including ``edge computing'', ``UAV'', ``drones'', ``benefits'', ``limitations'', and ``challenges''.
The Boolean operators \texttt{AND} and \texttt{OR} were used to combine search terms and broaden the scope of the search.

A search expression such as \texttt{"edge computing" AND ("UAV" OR "drone") AND ("benefits" OR "limitations" OR "challenges")} is expected to yield the most relevant literature for the research question.
However, this search expression is likely to focus on secondary literature such as surveys and existing literature reviews.
To shift the focus to primary literature, the search expression was simplified to \texttt{"edge computing" AND ("UAV" OR "drone")}.
The number of relevant publications in the mentioned databases for this simplified search term are shown in \cref{tab:databases}.
We note that some results may reflect duplicates.

\begin{table}
    \centering
    \caption{Number of Results for Our Search Term}
    \label{tab:databases}
    \begin{tabular}{ lr }
        \toprule
        \textbf{Database}    & \textbf{Results} \\
        \midrule
        Google Scholar       & 24 800           \\
        SpringerLink         & 3 429            \\
        ScienceDirect        & 1 528            \\
        \textbf{IEEE Xplore} & \textbf{869}     \\
        ACM Digital Library  & 373              \\
        \midrule
        \textbf{Total}       & \textbf{30,999}  \\
        \bottomrule
    \end{tabular}
\end{table}

This literature review will focus on an exhaustive review of the search results returned by the IEEE Xplore database, as the results are the most relevant to the research question and the amount allows for an exhaustive human review.

To ensure a comprehensive review, the search was not limited by publication date, allowing the inclusion of both foundational and recent works.
We only considered peer-reviewed articles written in English for inclusion.

\subsection{Study Selection}

After reviewing titles and abstracts, we selected the initial results of the search to remove irrelevant papers.
Next, we ensured the remaining articles directly addressed the research question by reviewing their full text and checking against our inclusion criteria.
Any articles that focused on a different topic or only tangentially mentioned UAVs or edge computing were excluded from the selection.

We ensured that only articles focusing on edge computing \emph{assisting} UAVs were included, while those treating UAVs \emph{as} edge servers were omitted.
For our selection, the edge server(s) must not be integrated within the UAV but be their own physical devices, requiring a wireless connection to the UAV for data transfer.
Integrating edge servers within the UAV is a valid approach, but overlooks the risks and limitations of wireless data transfer between UAVs and edge servers and is therefore not considered in this literature review.

\subsection{Data Extraction and Analysis}

For each included study, relevant data was extracted, such as study aim, methods, key findings, benefits and limitations identified, and proposed solutions or recommendations.
Most importantly, the role of the edge server in the system and the way of improving UAV performance were extracted.

Extracted data was synthesized and analyzed to identify common themes, contradictions, and gaps in the literature.
These findings formed the basis of the Results and Discussion sections.

\section{Results}
\label{sec:results}

\subsection{Search Outcomes}

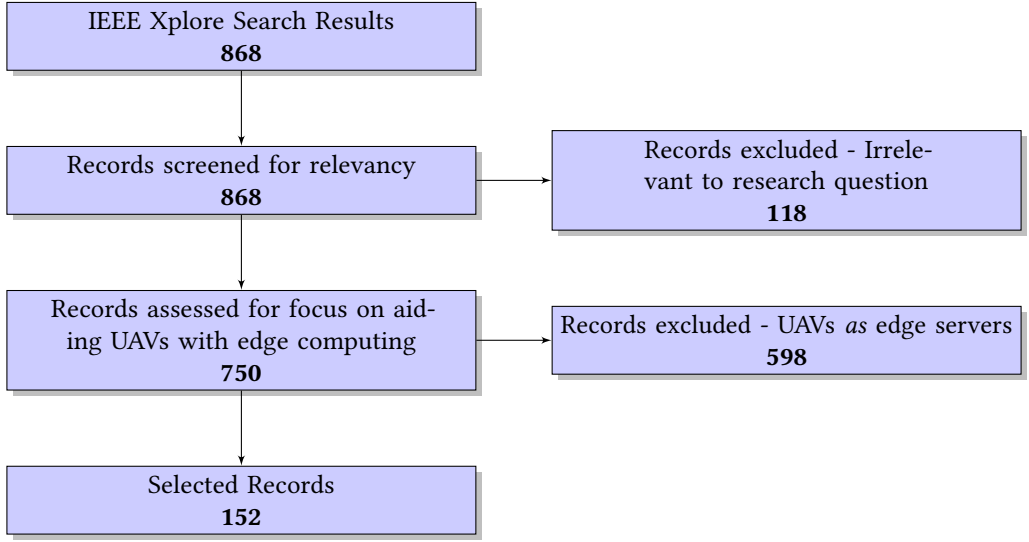
\begin{figure*}
    \centering
    \begin{tikzpicture}[node distance=1cm and 1cm, auto, every node/.style={align=center, text width=6cm}]

        \tikzstyle{block} = [rectangle, draw, fill=blue!20, minimum width=2em, minimum height=2em, text centered, drop shadow]
        \tikzstyle{line} = [draw, -latex']

        \node [block] (ieeexplore) {IEEE Xplore Search Results \\ \textbf{868}};

        \node [block, below=of ieeexplore] (screened) {Records screened for relevancy \\ \textbf{868}};
        \node [block, right=of screened] (excluded1) {Records excluded - Irrelevant to research question \\ \textbf{118}};

        \node [block, below=of screened] (assessed) {Records assessed for focus on aiding UAVs with edge computing \\ \textbf{750}};
        \node [block, right=of assessed] (excluded2) {Records excluded - UAVs \emph{as} edge servers \\ \textbf{598}};

        \node [block, below=of assessed] (selected) {Selected Records \\ \textbf{152}};

        \path [line] (ieeexplore) -- (screened);

        \path [line] (screened) -- (assessed);
        \path [line] (screened) -- (excluded1);

        \path [line] (assessed) -- (selected);
        \path [line] (assessed) -- (excluded2);

    \end{tikzpicture}
    \caption{Literature Selection Process}
    \label{fig:selection_process}
\end{figure*}

The results of database search and subsequent selection process are shown in \cref{fig:selection_process}.

Out of the 868 records returned by the IEEE Xplore search, 118 were excluded because they were not relevant to the research question.
The remaining 750 records were assessed for their focus on aiding UAVs with edge computing.
598 of these records were excluded because they focused on UAVs \emph{as} edge servers and not on edge servers \emph{assisting} UAVs.
The remaining 152 records were selected for the literature review.

This is an exhaustive search, leaving no article for the search term on IEEE Xplore unreviewed.
All identified literature was published in the last five years, from 2017 to 2023.
The records mainly come from the four different sources shown in \cref{tab:ieeesources}

\begin{table}
    \centering
    \caption{IEEE Xplore Literature Search Result Sources}
    \label{tab:ieeesources}
    \begin{tabular}{ l r }
        \toprule
        Source                     & Count        \\
        \midrule
        IEEE Conferences           & 409          \\
        IEEE Journals              & 325          \\
        IEEE Early Access Articles & 67           \\
        IEEE Magazines             & 57           \\
        Misc.
                                   & 10           \\
        \midrule
        \textbf{Total}             & \textbf{868} \\
        \bottomrule
    \end{tabular}
\end{table}

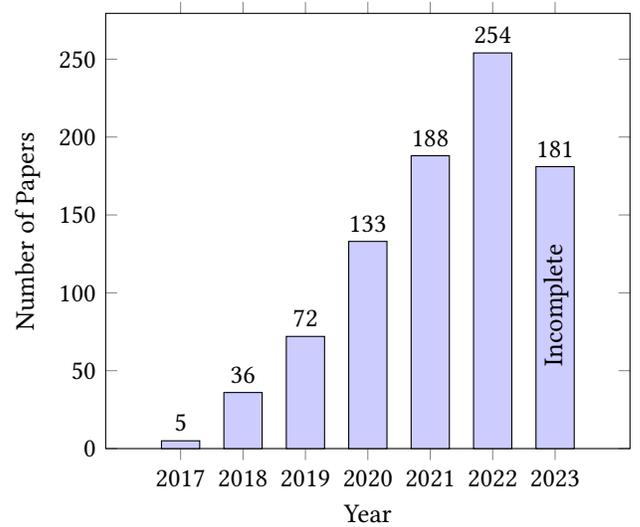
\begin{figure}
    \centering
    \resizebox{\linewidth}{!}{
        \begin{tikzpicture}
            \begin{axis}[
                    ybar,
                    ylabel={Number of Papers},
                    xlabel={Year},
                    symbolic x coords={2017, 2018, 2019, 2020, 2021, 2022, 2023},
                    xtick=data,
                    bar width=0.5cm,
                    enlarge x limits=0.2,
                    ymin=0,
                    nodes near coords,
                ]
                \addplot [fill=lightblue] coordinates {
                        (2017, 5)
                        (2018, 36)
                        (2019, 72)
                        (2020, 133)
                        (2021, 188)
                        (2022, 254)
                        (2023, 181)
                    } node[pos=1.0, above, yshift=-12ex, xshift=2ex, rotate=90] {Incomplete};
            \end{axis}
        \end{tikzpicture}
    }
    \caption{Yearly Distribution of Identified Papers}
    \label{fig:yearly_papers}
\end{figure}

As shown by the temporal distribution of published articles in \cref{fig:yearly_papers}, research in the area is accelerating.

The first paper found in the search was published in 2017, with a steady increase in the following years.
The number of papers published in 2023 is incomplete, as the data was collected in August 2023.
The earliest paper found in the search is by~\citet{Motlagh2017}, which outlines a use case for crowd surveillance using offloading for face detection and recognition.

\subsection{Common themes}

\begin{figure*}
    \centering
    \begin{tikzpicture}[node distance=0.5cm and -2.5cm, auto, every node/.style={align=center, text width=6cm}]

        \tikzstyle{block} = [rectangle, draw, fill=blue!20, minimum width=2em, minimum height=2em, text centered, drop shadow]
        \tikzstyle{line} = [draw, -latex']

        \node [block] (selected) {Selected Records \\ \textbf{152}};

        \node [block, below right=of selected] (theme1) {Offloading to edge server \\ \textbf{147}};
        \node [block, below right=of theme1] (subtheme1) {Vision capabilities \\ \textbf{75}};
        \node [block, below=of subtheme1] (subtheme2) {Trajectory planning \\ \textbf{47}};
        \node [block, below=of subtheme2] (subtheme3) {Artificial Intelligence/Applied Machine Learning \\ \textbf{47}};
        \node [block, below=of subtheme3] (subtheme4) {Energy efficient offloading \\ \textbf{34}};
        \node [block, below=of subtheme4] (subtheme5) {Object detection \\ \textbf{27}};
        \node [block, below=of subtheme5] (subtheme6) {Mapping \\ \textbf{7}};
        \node [block, below=of subtheme6] (subtheme7) {Face detection \\ \textbf{6}};
        \node [block, below=of subtheme7] (subtheme8) {Logistics \\ \textbf{5}};
        \node [block, below left=of subtheme8] (theme2) {Video transmission \\ \textbf{7}};

        \draw [line] (selected.south) |- (theme1.west);
        \draw [line] (selected.south) |- (theme2.west);
        \draw [line] (theme1.south) |- (subtheme1.west);
        \draw [line] (theme1.south) |- (subtheme2.west);
        \draw [line] (theme1.south) |- (subtheme3.west);
        \draw [line] (theme1.south) |- (subtheme4.west);
        \draw [line] (theme1.south) |- (subtheme5.west);
        \draw [line] (theme1.south) |- (subtheme6.west);
        \draw [line] (theme1.south) |- (subtheme7.west);
        \draw [line] (theme1.south) |- (subtheme8.west);
        \draw [line] (theme1.south) -- (theme2.north);

    \end{tikzpicture}
    \caption{Major Themes in Identified Literature}
    \label{fig:themes}
\end{figure*}
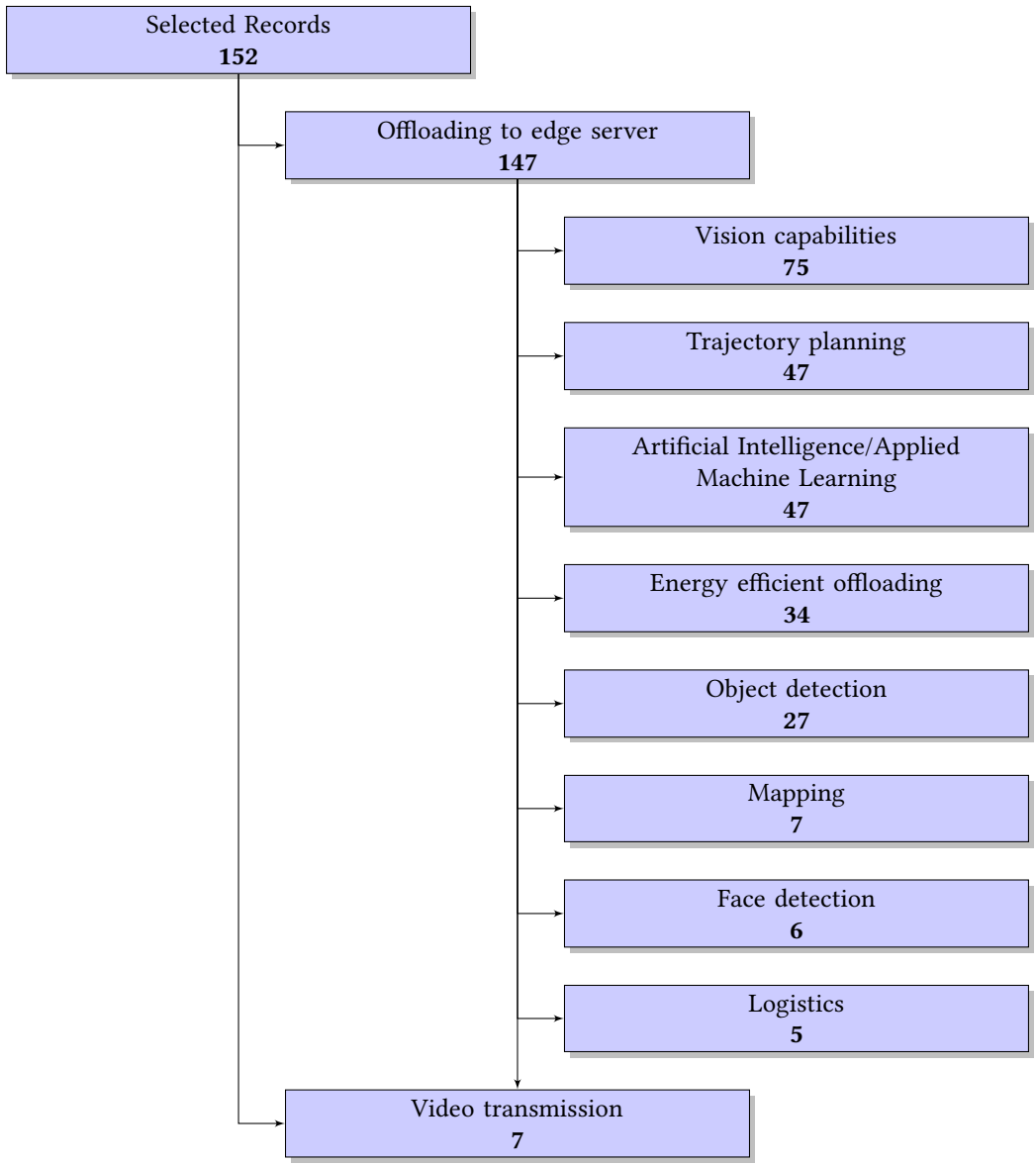

The selected records share common themes that are shown in \cref{fig:themes}, along with their absolute and relative frequency.
Common themes are not exclusive, a record can have multiple common themes -- therefore the sum of all absolute frequency counts is greater than the total number of selected records.
All but some records utilize offloading to an edge server, the remaining 5 records only use edge servers for video transmission and no other concrete computation.
A complete list of records can be found in \cref{sec:appendix}.

The listed common themes often encompass multiple applications.

\textbf{Offloading to edge servers} is utilized by all but 5 selected records.
Computationally expensive tasks can be moved to an edge server for multiple reasons: Typically, edge servers have increased computational power and faster, more reliable network conditions.
Aside from that edge servers are not limited by a battery, which on UAVs usually requires careful consideration on what tasks to execute and what to ignore.
The following themes are all applications that depend on or benefit from offloading to an edge server.

\textbf{Vision capabilities} describe all uses of cameras on the UAV to collect visible information.
This includes photos and videos, but also all applications that depend on cameras, such as industrial inspection, object and facial detection, photogrammetry and camera-based navigation.
\citet{Chen2021} use a camera to build a 3D map of the environment, which is then used to plan a better trajectory for the UAV.

\textbf{Trajectory planning} indicates (semi-)autonomous navigation, utilizing systems such as GPS or camera 3D information, combined with pathfinding to ensure a safe flight.
\citet{Seisa2022} offload sensor data to an edge server, which then returns an improved trajectory for the UAV to follow.

\textbf{Artificial Intelligence/Applied Machine Learning} incorporates various applications where ML inference is required for the UAV.
Typical ML models increase drastically in size and hardware requirements when ever-improving accuracy is expected.
For these use cases on-device computation is often not feasible because of computation/battery constraints, and it can be desirable to offload these tasks to dedicated edge servers.
There, large ML models can run unconstrained.
\citet{Li2021} utilize a large ML model running on an edge server to detect pine wilt disease in forests.

\textbf{Energy efficient offloading} highlights the previously mentioned concerns when UAV activities are limited by the battery.
In these cases, offloading tasks to an edge server can offer improved UAV runtime and increased task completion speed.
On the other hand, unnecessary offloading of computationally simple tasks can diminish UAV performance, when the optimal solution is to compute locally on the UAV and not send data to the server.
\citet{Cao2019} optimize the offloading of tasks to an edge server to maximize UAV runtime when inspecting wind farms.

\textbf{Object detection} often depends on the previously mentioned vision capabilities, but depending on the application other solutions such as Lidar are viable alternatives.
Often combined with ML models, objects can be reliably detected and tracked without human intervention.
Typical use cases are industrial inspection flights or surveillance tasks.
\citet{Yang2021} track vehicles using UAVs, offloading the detection and tracking to an edge server.

\textbf{Mapping} refers to generating 2D/3D map data from previously collected sensor data, such as camera recordings or Lidar measurements.
Accurate and detailed maps help with trajectory planning and can lead to more efficient flights.
\citet{Messous2020} discuss the use of UAVs for 3D mapping, offloading the sensor data to an edge server for processing.

\textbf{Face detection} is a similar application to object detection, but offers special use cases like identification and authentication.
Most face detection utilizes specific ML models to accurately identify faces.
\citet{Xu2022ispa} describe how an edge server can assist UAVs to detect and identify recipients of a package delivery.

\textbf{Logistics} describes the preparation and transportation of cargo using UAVs, which carry package to their destination.
This requires efficient trajectory planning and benefits from vision/ML advancements to allow for precise delivery.
\citet{Zhang2021} use UAVs to deliver packages to customers, assisted by an edge server to plan the optimal route.

\textbf{Video transmission} from the UAV is typically required for human pilots on the ground, allowing an operator to gain precise information about the location of the UAV.
Here, low latency and high throughput are essential.
\citet{Taleb2023} utilize an edge server to transmit video data from the UAV to the operator, allowing for precise control of the UAV.

\section{Discussion}
\label{sec:discussion}

\subsection{Summary of Major Findings}

The papers and articles returned by the scientific search engines outline two different, but interrelated use cases for combining UAVs and edge computing.

For one, UAV performance can be improved greatly through edge computing.
In this application, the focus is on using edge computing to enhance UAV functionalities, by offloading computational tasks from the UAV to edge servers, enhancing UAV efficiency and computational performance for on-device tasks.
This could be in the form of improved autonomous navigation~\cite{Chu2021}, mapping~\cite{Jharko2023}, real-time analysis of complex sensor data~\cite{Masuduzzaman2022}, and reduced energy consumption~\cite{Luo2023}.
An integral part of these systems are the edge servers, typically stationed on the ground.
As these servers are not battery-powered, unlike the UAV, they are able to complete computationally heavy or long-running tasks which would not be possible on the UAV.
This offloading to the edge allows the UAV to operate with reduced computational and energy demands, thus preserving battery life and allowing prolonged operations.

In contrast, most studies focus on the role of UAVs as mobile edge computing platforms themselves (e.g.,~\cite{Wang2021, Wu2022, Wang2020}).
In these scenarios, UAVs are used to carry and deploy edge computing capabilities to areas where they are needed.
The UAV can either run computational tasks directly on-device for optimal processing speed or act as a bridge between ground-based devices and edge servers positioned further away.
The decision whether to run computations directly on the UAV or on a ground station is mostly influenced by the processing capabilities required for the task and the limited battery capacity on the UAV.
In both cases, aerial coverage for devices on the ground is increased, even allowing for dynamic adjustments by repositioning the UAVs quickly.

This literature review focuses on the former, where UAVs are assisted by edge servers.
Statistically, there is four times as much research available on the second use case, where UAVs act as edge servers themselves (152 vs.
598 articles).
This makes the former use case interesting for a literature review, as potential research gaps are more likely to be identifiable.

Current literature focuses on a few main areas of edge-based assistance.
Most research suggests improvements to applications that were previously theoretically possible on UAVs, but had serious disadvantages compared to offloading them to an edge server.
Through increased wireless network speeds and reduced latencies, offloading often is the superior choice.

\subsection{Interpretation and Analysis}

The increasing interest in applications of UAVs follows the technological advancements in the area, making use cases viable that were previously out of reach for UAVs.
These advancements often stem from general improvements in computational power, which has been steadily increasing over the last decades~\cite{Mack2011}.
More precisely, increased computational power in UAVs and edge servers has allowed for real-time photo and video processing~\cite{Yang2020}, multiple types of visual detection and segmentation~\cite{Goudarzi2022}, more accurate machine learning applications~\cite{Hao2022} and 3D mapping and trajectory planning~\cite{Fukui2022} use cases.

The major area of improvement, found in every second selected article (75 of 152), is in visual processing, where camera quality has drastically increased over the last years.
Because of this, transmission and processing times have increased accordingly.
As most UAVs carry cameras for navigation purposes, new use cases require little to no hardware changes.
Complex tasks in this area include object~\cite{Wu2021}, pose, and face detection~\cite{Xu2022_2}, image classification and segmentation~\cite{Ilhan2021}, 2D/3D mapping~\cite{Shao2022} and obstacle avoidance.
These tasks typically utilize trained neural networks which often come with large memory requirements and noticeable processing times.
Typically, edge-assisted UAV systems offload the camera data -- photos and videos -- to the edge, where the trained model processes the data and optionally returns instructions to the UAV.
There is some research on smaller models running on the UAV, reducing the payload for transmission, but the practicality of this approach depends heavily on the exact model running on the UAV, UAV hardware specifications, and transmission speeds.

Another frequently researched area is in trajectory planning for UAVs.
This can have multiple benefits to UAV performance and there are different reasons as to why trajectory planning is researched.
If the UAV has to travel quickly across waypoints, e.g.
for logistic purposes~\cite{Xu2022ispa}, the necessity for efficient routes is obvious.
In other cases, when traversal speed is not the primary objective, there are other reasons for focusing on trajectory planning.
Assuming a UAV is being assisted by an edge server in completing a computational task, it is of importance to establish a reliable, high-throughput connection to the edge server.
As these servers are statically positioned, the UAV has to be in close proximity to it to establish this connection.
In these cases, trajectory planning is not primarily influenced by actual flight targets but by connection quality.
A UAV could move away from the shortest path if it allows for more offloading chances and thus increased task completion rates~\cite{Chen2021}.

Yet another prominent theme found in the literature is energy efficient offloading, which is caused by the limited UAV battery capacities.
When a UAV has to complete a computationally intensive task and there is an edge server nearby, offloading it preserves energy for the UAV, but possibly increases task completion time if the data transmission to the edge server is prolonged by poor signal strength or an overloaded edge server~\cite{Kim2019}.
While waiting for the edge server to complete the task, energy is wasted on staying in the air.
Because of this conflict, the mentioned factors need to be taken into careful consideration before deciding whether to offload tasks to edge servers~\cite{Deng2022}.

These themes form the major areas of research on edge-assisted UAV applications.
Influenced by commercially available UAVs, most research utilizes the typical hardware combination of a flight controller, wireless connectivity module, and cameras.
This can explain the focus on vision-based applications, given the ubiquity and ease-of-use of cameras on UAVs.

\subsection{Research Gaps}

Upon exhaustive analysis of the existing literature, several research gaps highlight areas requiring further exploration:

\paragraph*{Reliance on Future Network Standards:}
A significant portion of the current research concentrates on utilizing futuristic networking standards such as 5G, 6G, and beyond (e.g.,~\cite{Hayat2021, Bera2022}).
However, there's a notable scarcity of studies that study the practical implications of integrating UAVs with current, widespread technologies like 4G.
As newer standards may take several years to become universally adopted, understanding UAV operations in the context of existing networks remains crucial.

\paragraph*{Theoretical Dominance:}
The bulk of the existing literature leans towards theoretical models and simulations.
Research that examines real-world scenarios -- such as the impact of weather conditions, wind variations, day-night transitions, and real wireless networking challenges -- is rare.

\paragraph*{Urban and Short-Term Bias:}
Much of the research is biased towards urban environments and short-term UAV deployments.
Studies focusing on longer flight durations, such as those akin to gliders, or operations over vast, less-populated terrains, are not as prevalent.
Notably, UAVs become particularly compelling in non-urban environments, where traditional static infrastructure is often lacking or non-existent.
The capability of UAVs to serve these underrepresented areas emphasizes their value and potential.

\paragraph*{UAV Safety Concerns:}
While there's an understandable enthusiasm surrounding the innovative applications of UAVs and their information security, there appears to be an oversight when it comes to physical safety.
Specifically, strategies to prevent UAV collisions with obstacles, other drones, or crucially, ensuring safe behavior during connection losses, are not comprehensively addressed.

\paragraph*{Risk Analysis and Recovery:}
The available research seems to lack comprehensive studies focusing on UAV fail rates, their common causes, and subsequent recovery mechanisms.
Understanding typical UAV failure scenarios, their probabilities, and formulating strategies to swiftly and safely address possible accidents, is crucial to ensure robust and reliable UAV deployments.

\paragraph*{Economic Feasibility:}
While the technical dimensions of UAV-Edge Computing integrations are often explored, there's a lack of comprehensive studies on the economic aspects.
Questions related to the cost-benefit analysis and long-term financial viability of deploying edge-assisted UAV systems remain largely unanswered.

\paragraph*{Interdisciplinary Synergies:}
There's potential to explore synergies between UAV-Edge Computing and other fields, such as renewable energy (for prolonged UAV operations), advanced materials (for UAV design enhancements), or behavioral sciences (to understand human-UAV interactions in public spaces).

The relative underrepresentation of UAVs utilizing ground-based edge servers in the literature highlights the need for a more balanced research outlook, especially given the potential advantages this synergy promises.
Addressing the listed research gaps could not only lead to a more comprehensive understanding of the UAV and edge computing domain but also allow for improvements that are both technically sound and practically viable.
\section{Limitations}
\label{sec:limitations}

Several limitations are inherent in our systematic literature review, which readers should bear in mind:

\paragraph*{Scope of Search Engines:} We relied on one specific scientific search engine -- IEEE Xplore -- to obtain the research papers.
As a result, limitations specific to IEEE Xplore, such as its collection bias or specific indexing criteria, directly influence the scope and breadth of our review.
Moreover, relevant articles from other sources or less popular databases might not be included in our review.

\paragraph*{Theoretical Versus Experimental Findings:} A significant proportion of the reviewed studies predominantly emphasize theoretical analyses and models.
In contrast, empirical studies that present real-life experimental results or field tests remain relatively rare.

\paragraph*{Publication Bias:} There's an inherent publication bias in most academic fields where studies with positive or significant results are more likely to get published than those with neutral or negative findings~\cite{Song2010}.
Most frequently researched in biomedical fields, this bias affects other domains as well.
This bias could have influenced the themes and results we presented.

\paragraph*{Time Constraint:} Due to the dynamic nature of research, newer studies may have been published after our review's cutoff date.
As a result, our findings might not contain the very latest trends or breakthroughs in the intersection of UAVs and edge computing and the label ``exhaustive review'' might not be appropriate anymore.

\paragraph*{Language Bias:} Our review was limited to articles published in English.
Important findings or developments presented in articles in other languages might not be included.

\paragraph*{Selection Criteria and Search Terms:} While we employed strict criteria for the selection of papers, the choice of our search terms may inherently possess biases or shortcomings.
The possibility exists that relevant articles using synonyms or alternative phrasings to our chosen terms were overlooked.
Consequently, some important papers might have been inadvertently excluded, while others of lesser relevance were included.
Additionally, our focus on UAVs assisted by edge servers might have limited the breadth of insights in the broader domain of UAVs and edge computing.

\paragraph*{Interpretative Nature:} Systematic literature reviews involve a degree of interpretation when categorizing, analyzing, and synthesizing research findings.
Different researchers might interpret or prioritize findings differently, potentially leading to different conclusions or emphases.

We acknowledge these limitations and have aimed to be as comprehensive, unbiased, and rigorous as possible in our approach.
Future reviews might consider addressing some of these limitations by broadening the scope, considering different databases and multi-language studies, or employing different methodologies for data extraction and synthesis.

\section{Conclusion}
\label{sec:conclusion}

The combination of UAV technology and edge computing has allowed for new advancements in aerial computation and service delivery.
Through this literature review, two primary paradigms emerged: edge servers assisting UAVs to enhance their efficiency and computational power, and UAVs acting as mobile edge servers to serve ground-based devices.
Interestingly, while the former is less prevalent in the current literature, it offers a promising area for potential research opportunities, especially given the rapid evolution of wireless networking and computational capabilities.

One of the main drivers behind these advancements is the continuous growth in computational power, making previously challenging tasks feasible for UAVs.
Real-time visual processing, comprehensive visual detection, and state-of-the-art machine learning applications are no longer confined to resource-rich environments but are being brought to aerial devices.
With improvements in camera quality, visual processing stands out as a dominant research theme, allowing for innovative applications in object detection, image segmentation, and aerial mapping, among others.

Trajectory planning, while traditionally used exclusively for efficient navigation, can offer new approaches to UAV edge assistance.
The importance of ensuring high-quality connections to static edge servers can force UAVs to prioritize connection strength over direct navigational efficiency.
This requires a complicated balance between optimal routes, offloading opportunities, and task completion rates.

The inherent limitation of UAV battery life also highlights the need for efficient offloading strategies.
While edge servers can indeed reduce the computational burden on UAVs, ensuring that offloading decisions are made effectively, considering both energy and time constraints, remains a challenge.

In retrospect, the combination of UAVs and edge computing is an evolving area of research, driven by both technological advancements and new application demands.
The current trajectory of research, influenced heavily by commercially available UAV hardware, is leading towards a more connected and efficient aerial ecosystem.
As this domain continues to grow, it will be valuable for future research to explore lesser-studied areas, addressing the current gaps, and further understanding how UAVs and edge computing can best complement each other.

\begin{acks}
    Funded by the \grantsponsor{BMBF}{Bundesministerium für Bildung und Forschung (BMBF, German Federal Ministry of Education and Research)}{https://www.bmbf.de/bmbf/en} -- \grantnum{BMBF}{16KISK183}.
\end{acks}

\balance
\bibliographystyle{ACM-Reference-Format}
\bibliography{bibliography.bib}

\appendix
\onecolumn\section{List of Reviewed Publications}
\label{sec:appendix}

We provide a list of all identified literature in \cref{tab:literature}.
We use acronyms for common themes as shown in \cref{tab:acronyms}.

\tiny
\begin{longtable}[c]{ l l }
    \caption{Selected Literature}
    \label{tab:literature}
    \\
    \toprule
    Publication                                                                                                                                                                & Themes             \\
    \midrule
    \endfirsthead

    \toprule
    Publication                                                                                                                                                                & Themes             \\
    \midrule
    \endhead

    \bottomrule
    \endfoot

    \bottomrule
    \endlastfoot

    An Evaluation of Edge Computing Platform for Reliable Automated Drones~\cite{Yoshimoto2020}                                                                                & OE                 \\
    A Novel Optimization Strategy For Computation Offloading in the UAV-assisted Edge Computing ~\cite{Vijaybhai2022}                                                          & OE                 \\
    Rate Splitting on Mobile Edge Computing for UAV-Aided IoT Systems~\cite{Han2020}                                                                                           & OE                 \\
    Artificial Intelligence Empowered UAVs Data Offloading in Mobile Edge Computing~\cite{Fragkos2020}                                                                         & OE                 \\
    Task-Oriented Multi-Modal Communication Based on Cloud-Edge-UAV Collaboration~\cite{Ren2023}                                                                               & OE                 \\
    IoT-based CO2 Gas-level Monitoring and Automated Decision-making System in Smart Factory using UAV-assisted MEC~\cite{Masuduzzaman2022}                                    & OE                 \\
    Reducing the Mission Time of Drone Applications through Location-Aware Edge Computing~\cite{Kasidakis2021}                                                                 & OE                 \\
    Let's Trade in the Future! A Futures-Enabled Fast Resource Trading Mechanism in Edge Computing-Assisted UAV Networks~\cite{Liwang2021}                                     & OE                 \\
    Online Computation Offloading and Traffic Routing for UAV Swarms in Edge-Cloud Computing~\cite{Liu2020}                                                                    & OE                 \\
    Resource Inference for Task Migration in Challenged Edge Networks with RITMO~\cite{Sacco2020}                                                                              & OE                 \\
    A Self-Learning Strategy for Task Offloading in UAV Networks~\cite{Sacco2022}                                                                                              & OE                 \\
    Sustainable Task Offloading in UAV Networks via Multi-Agent Reinforcement Learning~\cite{Sacco2021}                                                                        & OE                 \\
    Secure Task Offloading for MEC-Aided-UAV System~\cite{Chen2023}                                                                                                            & OE                 \\
    Planning Computation Offloading on Shared Edge Infrastructure for Multiple Drones~\cite{Polychronis2022}                                                                   & OE                 \\
    Location-Based Beamforming Architecture for Efficient Farming Applications with Drones~\cite{Wang2019}                                                                     & OE                 \\
    Software-defined network based resource allocation in distributed servers for unmanned aerial vehicles~\cite{Shukla2018}                                                   & OE                 \\
    Utility Based Scheduling for Multi-UAV Search Systems in Disaster-Hit Areas~\cite{Miyano2019}                                                                              & OE                 \\
    Power-Delay Trade-off for Heterogenous Cloud Enabled Multi-UAV Systems~\cite{Duan2019}                                                                                     & OE                 \\
    Cloud-Fog-Edge Computing in Smart Agriculture in the Era of Drones: a systematic survey~\cite{Dhifaoui2022}                                                                & OE                 \\
    Secure Data Sharing in UAV-assisted Crowdsensing: Integration of Blockchain and Reputation Incentive~\cite{Xie2021}                                                        & OE                 \\
    Delay-Optimal Task Offloading for UAV-Enabled Edge-Cloud Computing Systems~\cite{Almutairi2022}                                                                            & OE                 \\
    Stochastic Coded Offloading Scheme for Unmanned-Aerial-Vehicle-Assisted Edge Computing~\cite{Ng2023}                                                                       & OE                 \\
    An Intelligent Task Offloading Algorithm (iTOA) for UAV Network~\cite{Chen2019}                                                                                            & OE                 \\
    Toward Performance Efficient UAV Task Scheduling in Cloud Native Edge~\cite{Huang2022}                                                                                     & OE                 \\
    Deep Reinforcement Learning-Based Resource Allocation for UAV-Enabled Federated Edge Learning~\cite{Liu2023}                                                               & OE, AI             \\
    Learning-based Multi-Drone Network Edge Orchestration for Video Analytics~\cite{Qu2022}                                                                                    & OE, AI, VC         \\
    Edge Computing in IoT Ecosystems for UAV-Enabled Early Fire Detection~\cite{Kalatzis2018}                                                                                  & OE, AI, VC         \\
    Fractus: Orchestration of Distributed Applications in the Drone-Edge-Cloud Continuum~\cite{Grigoropoulos2022}                                                              & OE, AI, VC         \\
    Edge Computing Based Abnormal Behavior Learning for Mental Disorder Detection Through UAV Surveillance~\cite{Hao2022}                                                      & OE, AI, VC         \\
    Drone Aided Thermal Mapping for Selective Irrigation of Localized Dry Spots~\cite{Jalajamony2023}                                                                          & OE, AI, VC         \\
    UAV Swarms in Smart Agriculture: Experiences and Opportunities~\cite{Qu2022escience}                                                                                       & OE, AI, VC         \\
    A YOLOv3-based Learning Strategy for Real-time UAV-based Forest Fire Detection~\cite{Jiao2020}                                                                             & OE, AI, VC         \\
    Offloading Deep Learning Powered Vision Tasks From UAV to 5G Edge Server With Denoising~\cite{Ozer2023}                                                                    & OE, AI, VC         \\
    Edge4Sys: A Device-Edge Collaborative Framework for MEC based Smart Systems~\cite{9285667}                                                                                 & OE, AI, VC, FD     \\
    Edge4FR: A Novel Device-Edge Collaborative Framework for Facial Recognition in Smart UAV Delivery Systems~\cite{Xu2022ccis}                                                & OE, AI, VC, FD     \\
    Proof-of-Concept of Uncompressed 4K Video Transmission from Drone through mmWave~\cite{Takaku2020}                                                                         & OE, AI, VC, FD     \\
    EdgeDuet: Tiling Small Object Detection for Edge Assisted Autonomous Mobile Vision~\cite{Yang2023}                                                                         & OE, AI, VC, OD     \\
    Edge-based Realtime Image Object Detection for UAV Missions~\cite{Wu2021}                                                                                                  & OE, AI, VC, OD     \\
    DyCOCo: A Dynamic Computation Offloading and Control Framework for Drone Video Analytics~\cite{Qu2019}                                                                     & OE, AI, VC, OD     \\
    Bandwidth-Efficient Live Video Analytics for Drones Via Edge Computing~\cite{Wang2018}                                                                                     & OE, AI, VC, OD     \\
    Optimal Offloading of Computing-intensive Tasks for Edge-aided Maritime UAV Systems~\cite{Li2022vtc}                                                                       & OE, AI, VC, OD     \\
    Blockchain-Assisted UAV-Employed Casualty Detection Scheme in Search and Rescue Mission in the Internet of Battlefield Things~\cite{Masuduzzaman2020}                      & OE, AI, VC, OD     \\
    Seamless Virtualized Controller Migration for Drone Applications~\cite{An2019}                                                                                             & OE, AI, VC, OD     \\
    PA-Offload: Performability-Aware Adaptive Fog Offloading for Drone Image Processing~\cite{Machida2021}                                                                     & OE, AI, VC, OD     \\
    A Remote Sensing and Airborne Edge-Computing Based Detection System for Pine Wilt Disease~\cite{Li2021}                                                                    & OE, AI, VC, OD     \\
    Offloading Deep Learning Empowered Image Segmentation from UAV to Edge Server~\cite{Ilhan2021}                                                                             & OE, AI, VC, OD     \\
    Edge Computational Offloading for Corrosion Inspection in Industry 4.0~\cite{ KhadmaouiBichouna2022}                                                                       & OE, AI, VC, OD     \\
    Edge-Cloud Architectures Using UAVs Dedicated To Industrial IoT Monitoring And Control Applications~\cite{Salhaoui2020}                                                    & OE, AI, VC, OD     \\
    Task-Oriented Image Transmission for Scene Classification in Unmanned Aerial Systems~\cite{Kang2022}                                                                       & OE, AI, VC, OD     \\
    Application of Unmanned Aerial Vehicles in Smart Cities using Computer Vision Techniques~\cite{Shirazi2020}                                                                & OE, AI, VC, OD     \\
    Offloading Optimization in Edge Computing for Deep-Learning-Enabled Target Tracking by Internet of UAVs~\cite{Yang2021}                                                    & OE, AI, VC, OD, TP \\
    Towards a Mobile App Platform for Personalized UAV Fleets using Edge and Cloud~\cite{Raj2023}                                                                              & OE, AI, VC, OD, TP \\
    CADET: Control-Aware Dynamic Edge Computing for Real-Time Target Tracking in UAV Systems~\cite{FlorenzanReyes2023}                                                         & OE, AI, VC, OD, TP \\
    Intelligent Search and Find System for Robotic Platform Based on Smart Edge Computing Service~\cite{Barnawi2020}                                                           & OE, AI, VC, OD, TP \\
    The fixed-point landing of the quad-rotor UAV based on YOLOv5~\cite{Li2022cac}                                                                                             & OE, AI, VC, OD, TP \\
    An Edge-Controlled Outdoor Autonomous UAV for Colorwise Safety Helmet Detection and Counting of Workers in Construction Sites~\cite{Sharma2021}                            & OE, AI, VC, OD, TP \\
    A Novel Adaptive Computation Offloading Strategy for Collaborative DNN Inference over Edge Devices~\cite{Zhang2022}                                                        & OE, AI, VC, OD, TP \\
    Intelli-Eye: An UAV Tracking System with Optimized Machine Learning Tasks Offloading~\cite{Yang2019}                                                                       & OE, AI, VC, OD, TP \\
    Real-time Crop Classification Using Edge Computing and Deep Learning~\cite{Yang2020}                                                                                       & OE, AI, VC, TP     \\
    UAV-enabled Edge Computing for Optimal Task Distribution in Target Tracking~\cite{Goudarzi2022}                                                                            & OE, AI, VC, TP     \\
    Edge-Based Live Video Analytics for Drones~\cite{Wang2019mic}                                                                                                              & OE, AI, VC, VT     \\
    Implementation and Investigation of High Endurance UAVs in a 4G LTE Network for Monitoring Power Lines~\cite{Ahmed2023}                                                    & OE, AI, VC, VT     \\
    Policy-Based Function-Centric Computation Offloading for Real-Time Drone Video Analytics~\cite{Chemodanov2019}                                                             & OE, AI, VC, VT     \\
    Research on Intelligent Mobile Edge Computing and Task Unloading Method of UAV~\cite{Zheng2023}                                                                            & OE, EE             \\
    Optimal Task-UAV-Edge Matching for Computation Offloading in UAV Assisted Mobile Edge Computing~\cite{Kim2019}                                                             & OE, EE             \\
    Energy Efficient UAV-Based Service Offloading Over Cloud-Fog Architectures~\cite{Alharbi2022}                                                                              & OE, EE             \\
    Energy Efficient Resource Allocation and Computation Offloading Strategy in a UAV-enabled Secure Edge-Cloud Computing System~\cite{Khan2020}                               & OE, EE             \\
    UAV-Aided Energy-Efficient Edge Computing Networks: Security Offloading Optimization ~\cite{Gu2022}                                                                        & OE, EE             \\
    Distributed Location-Aware Task Offloading in Multi-UAVs Enabled Edge Computing~\cite{Liu2022}                                                                             & OE, EE             \\
    Task Offloading in Multi-Access Edge Computing Enabled UAV-Aided Emergency Response Operations~\cite{Akter2023}                                                            & OE, EE             \\
    Real-time Data Acquisition and Processing under Mobile Edge Computing-assisted UAV System~\cite{Zeng2022}                                                                  & OE, EE             \\
    Energy optimization for Cellular-Connected UAV Mobile Edge Computing Systems~\cite{Wu2021}                                                                                 & OE, EE             \\
    Spectrum-Aware Mobile Edge Computing for UAVs Using Reinforcement Learning~\cite{9708942}                                                                                  & OE, EE             \\
    FEEL-enhanced Edge Computing in Energy Constrained UAV-aided IoT Networks~\cite{Sharma2023}                                                                                & OE, EE             \\
    Energy Optimization for Cellular-Connected Multi-UAV Mobile Edge Computing Systems with Multi-Access Schemes~\cite{Hua2018}                                                & OE, EE             \\
    Offloading Optimization for Energy-Minimization Secure UAV-Edge-Computing Systems~\cite{Gu2021}                                                                            & OE, EE             \\
    Intelligent Computation Offloading Mechanism of UAV in Edge Computing~\cite{Kasidakis2021}                                                                                 & OE, EE             \\
    Cellular-Connected Multi-UAV MEC Networks: An Online Stochastic Optimization Approach~\cite{Xu2022tcomm}                                                                   & OE, EE             \\
    A Coded Distributed Computing Framework for Task Offloading from Multi-UAV to Edge Servers~\cite{Guo2021}                                                                  & OE, EE             \\
    Optimal Edge Computing for Infrastructure-Assisted UAV Systems~\cite{Callegaro2021}                                                                                        & OE, EE             \\
    Stochastic Resource Management and Trajectory Optimization for Cellular-Connected Multi-UAV Mobile Edge Computing Systems~\cite{Xu2022isncc}                               & OE, EE             \\
    Energy Minimization for MEC-enabled Cellular-Connected UAV: Trajectory Optimization and Resource Scheduling~\cite{Lv2020}                                                  & OE, EE             \\
    Optimal Computation Offloading in Edge-Assisted UAV Systems~\cite{Callegaro2018}                                                                                           & OE, EE             \\
    Resource Inference for Sustainable and Responsive Task Offloading in Challenged Edge Networks~\cite{Sacco2021tgcn}                                                         & OE, EE             \\
    Cooperative Computation Offloading for UAVs: A Joint Radio and Computing Resource Allocation Approach~\cite{Zhu2018}                                                       & OE, EE             \\
    A Game Theory Based Efficient Computation Offloading in an UAV Network~\cite{Messous2019}                                                                                  & OE, EE             \\
    Machine Learning Based Edge-Assisted UAV Computation Offloading for Data Analyzing~\cite{Kim2020}                                                                          & OE, EE, AI         \\
    Energy-Efficient Task Offloading of Edge-Aided Maritime UAV Systems~\cite{Li2023}                                                                                          & OE, EE, AI, VC, OD \\
    Edge Computing in 5G for Drone Navigation: What to Offload?~\cite{Hayat2021}                                                                                               & OE, EE, AI, VC, OD \\
    A Holistic Service Provision Strategy for Drone-as-a-Service in MEC-based UAV Delivery~\cite{Chu2021}                                                                      & OE, EE, LO         \\
    Edge Computing Enabled Energy-Efficient Multi-UAV Cooperative Target Search~\cite{Luo2023edge}                                                                             & OE, EE, TP         \\
    Energy-Efficient UAV-Aided Target Tracking Systems Based on Edge Computing~\cite{Deng2022}                                                                                 & OE, EE, VC         \\
    Energy-Efficient Computation Offloading for Secure UAV-Edge-Computing Systems~\cite{Bai2019}                                                                               & OE, EE, VC         \\
    Computation offloading game for an UAV network in mobile edge computing~\cite{Messous2017}                                                                                 & OE, EE, VC         \\
    Delay and Energy Consumption Oriented UAV Inspection Business Collaboration Computing Mechanism in Edge Computing Based Electric Power IoT~\cite{Shao2023}                 & OE, EE, VC         \\
    MEC-Driven UAV Routine Inspection System in Wind Farm under Wind Influence~\cite{Cao2019}                                                                                  & OE, EE, VC         \\
    Low Energy Consumption Power Inspection Solution Based on Cloud-Fog-Edge Ladder Collaboration Technology~\cite{Zhao2022}                                                   & OE, EE, VC         \\
    Efficient Face Recognition via Multi-UAV-Edge Collaboration in UAV Delivery Service~\cite{Xu2022ispa}                                                                      & OE, LO, FD, AI, VC \\
    EXPRESS: An Energy-Efficient and Secure Framework for Mobile Edge Computing and Blockchain based Smart Systems~\cite{9286009}                                              & OE, LO, FD, AI, VC \\
    An Edge based Federated Learning Framework for Person Re-identification in UAV Delivery Service~\cite{Zhang2021}                                                           & OE, LO, FD, AI, VC \\
    A Novel Security Framework for Edge Computing based UAV Delivery System~\cite{Yao2021}                                                                                     & OE, LO, TP, AI, VC \\
    UAV-Aided Ultra-Reliable Low-Latency Computation Offloading in Future IoT Networks~\cite{Haber2021}                                                                        & OE, VC             \\
    Robust Edge Computing in UAV Systems via Scalable Computing and Cooperative Computing~\cite{Liu2021}                                                                       & OE, VC             \\
    Energy Minimization for Cellular-Connected UAV-MEC Patrol Inspection Systems~\cite{Wang2022}                                                                               & OE, VC             \\
    MARbLE: Multi-Agent Reinforcement Learning at the Edge for Digital Agriculture~\cite{Boubin2022}                                                                           & OE, VC             \\
    Drone-Aided Detection of Weeds: Transfer Learning for Embedded Image Processing~\cite{Koshelev2023}                                                                        & OE, VC             \\
    SeReMAS: Self-Resilient Mobile Autonomous Systems Through Predictive Edge Computing~\cite{Callegaro2021seremas}                                                            & OE, VC             \\
    3D Object Detection for Aerial Platforms via Edge Computing: An Experimental Evaluation~\cite{Lianides2022}                                                                & OE, VC             \\
    Semi-Autonomous Industrial Robotic Inspection: Remote Methane Detection in Oilfield~\cite{Filho2018}                                                                       & OE, VC             \\
    Real-Time Change Detection At the Edge~\cite{Gadiraju2022}                                                                                                                 & OE, VC             \\
    A Cloud Edge Collaborative Intelligence Method of Insulator String Defect Detection for Power IIoT~\cite{Song2021}                                                         & OE, VC             \\
    EagleEYE: Aerial Edge-enabled Disaster Relief Response System~\cite{Ardiansyah2020}                                                                                        & OE, VC             \\
    APRON: an Architecture for Adaptive Task Planning of Internet of Things in Challenged Edge Networks~\cite{Ventrella2019}                                                   & OE, VC             \\
    DEEP: A Vertical-Oriented Intelligent and Automated Platform for the Edge and Fog~\cite{Guimaraes2021}                                                                     & OE, VC, AI, TP, OD \\
    Edge Computing Assisted Autonomous Flight for UAV: Synergies between Vision and Communications~\cite{Chen2021}                                                             & OE, VC, MP TP      \\
    Deep Reinforcement Learning Based Computation Offloading and Trajectory Planning for Multi-UAV Cooperative Target Search~\cite{Luo2023}                                    & OE, VC, OD         \\
    UAV-Based IoT Platform: A Crowd Surveillance Use Case~\cite{Motlagh2017}                                                                                                   & OE, VC, OD         \\
    Edge-Assisted Intelligent Video Compression for Live Aerial Streaming~\cite{Liu2022tgcn}                                                                                   & OE, VC, TP         \\
    Towards beyond Visual Line of Sight Piloting of UAVs with Ultra Reliable Low Latency Communication~\cite{Ozger2018}                                                        & OE, VC, TP         \\
    Empowering Industry 4.0 and Autonomous Drone Scouting use cases through 5G-DIVE Solution~\cite{Conceicao2021}                                                              & OE, VC, TP         \\
    Dynamic Distributed Computing for Infrastructure-Assisted Autonomous UAVs~\cite{Callegaro2020}                                                                             & OE, VC, TP         \\
    Robot/UAV Indoor Visual SLAM in Smart Cities Based on Remote Data Processing~\cite{Jharko2023}                                                                             & OE, VC, TP, MP     \\
    Accurate and Efficient Multi-robot Collaborative Stereo SLAM for Mars Exploration~\cite{Shao2022}                                                                          & OE, VC, TP, MP     \\
    Vineyard Digital Twin: construction and characterization via UAV images -- DIWINE Proof of Concept~\cite{Edemetti2022}                                                     & OE, VC, TP, MP     \\
    A Map Building and Sharing Framework for Multiple UAV Systems~\cite{Silva2022}                                                                                             & OE, VC, TP, MP     \\
    IoT Edge Server ROS Node Allocation Method for Multi-SLAM on Many-core~\cite{Fukui2022}                                                                                    & OE, VC, TP, MP     \\
    Reconfigurable Intelligence Surface Aided UAV-MEC Systems With NOMA~\cite{Xu2022lcomm}                                                                                     & OE, TP             \\
    Mobile Edge Computing for Cellular-Connected UAV: Computation Offloading and Trajectory Optimization~\cite{Cao2018}                                                        & OE, TP             \\
    MEC-assisted Dynamic Geofencing for 5G-enabled UAV~\cite{Bera2022}                                                                                                         & OE, TP             \\
    Joint Computation Offloading and Routing Optimization for UAV-Edge-Cloud Computing Environments~\cite{Liu2018}                                                             & OE, TP             \\
    An Edge Architecture Oriented Model Predictive Control Scheme for an Autonomous UAV Mission~\cite{Seisa2022}                                                               & OE, TP             \\
    Toward 5G Edge Computing for Enabling Autonomous Aerial Vehicles~\cite{ Damigos2023}                                                                                       & OE, TP             \\
    PACED-5G: Predictive Autonomous Control using Edge for Drones over 5G~\cite{Sankaranarayanan2023}                                                                          & OE, TP             \\
    On the Feasibility of Infrastructure Assistance to Autonomous UAV Systems~\cite{Baidya2020}                                                                                & OE, TP             \\
    AdaDrone: Quality of Navigation Based Neural Adaptive Scheduling for Edge-Assisted Drones~\cite{Chen2022}                                                                  & OE, TP             \\
    Using 5G to Bring More than just Bits to Homes~\cite{Tan2020}                                                                                                              & OE, TP             \\
    UAV mission optimization in 5G: On reducing MEC service relocation~\cite{SiMohammed2020}                                                                                   & OE, TP             \\
    Fine-Grained Task Offloading for UAV via MEC-Enabled Networks~\cite{Huang2019}                                                                                             & OE, TP             \\
    UAV Cluster-Based Video Surveillance System Optimization in Heterogeneous Communication of Smart Cities~\cite{Jin2020}                                                     & OE, TP             \\
    UAVs Traffic Control Based on Multi-Access Edge Computing~\cite{Bekkouche2018}                                                                                             & OE, TP             \\
    A Kubernetes-Based Edge Architecture for Controlling the Trajectory of a Resource-Constrained Aerial Robot by Enabling Model Predictive Control~\cite{Seisa2022predictive} & OE, TP             \\
    A3D: Adaptive, Accurate, and Autonomous Navigation for Edge-Assisted Drones~\cite{Zeng2023}                                                                                & OE, TP             \\
    MEC-assisted Low Latency Communication for Autonomous Flight Control of 5G-Connected UAV~\cite{Solanki2023}                                                                & OE, TP             \\
    A Traffic-Aware Approach for Enabling Unmanned Aerial Vehicles (UAVs) in Smart City Scenarios~\cite{ElSayed2019}                                                           & OE, TP             \\
    Edge Cloud Resource-aware Flight Planning for Unmanned Aerial Vehicles~\cite{Bekkouche2019}                                                                                & OE, TP             \\
    Virtualized Control Over Fog: Interplay Between Reliability and Latency~\cite{Inaltekin2018}                                                                               & OE, TP             \\
    Edge Computing for Visual Navigation and Mapping in a UAV Network~\cite{Messous2020}                                                                                       & OE, TP, MP         \\
    Dynamic Online Trajectory Planning for a UAV-Enabled Data Collection System~\cite{Li2022tvt}                                                                               & TP                 \\
    Quality of Service Assessment of Live Video Streaming with a Remote-Controlled Drone~\cite{Loh2018}                                                                        & VT                 \\
    Dynamic cloud service placement for live video streaming with a remote-controlled drone~\cite{Wamser2017}                                                                  & VT                 \\
    Locomotion-Based UAV Control Toward the Internet of Senses~\cite{Sehad2023}                                                                                                & VT, TP             \\
    VR-Based Immersive Service Management in B5G Mobile Systems: A UAV Command and Control Use Case~\cite{Taleb2023}                                                           & VT, TP             \\
\end{longtable}
\normalsize

\begin{table}
    \centering
    \caption{Acronyms Used in Common Themes}
    \label{tab:acronyms}
    \begin{tabular}{ll}
        \toprule
        \textbf{Acronym} & \textbf{Common Theme}                            \\
        \midrule
        AI               & Artificial Intelligence/Applied Machine Learning \\
        EE               & Energy efficient offloading                      \\
        FD               & Face detection                                   \\
        LO               & Logistics                                        \\
        MP               & Mapping                                          \\
        OD               & Object detection                                 \\
        OE               & Offloading to edge server                        \\
        TP               & Trajectory planning                              \\
        VC               & Vision capabilities                              \\
        VT               & Video transmission                               \\
        \bottomrule
    \end{tabular}
\end{table}

\end{document}